\documentclass{article}

\usepackage{arxiv}
\usepackage{graphicx}
\usepackage{booktabs}
\usepackage{float}
\usepackage{amsmath}
\usepackage{hyperref}
\usepackage{array}
\usepackage{longtable}

\title{Security Hardening Using FABRIC: Implementing a Unified Compliance Aggregator for Linux Servers}

\author{
    Sheldon Paul \\
    Department of Computational, Engineering, and Mathematical Sciences\\
    Texas A\&M University- San Antonio\\
    San Antonio, Texas \\
    \texttt{sthil01@jaguar.tamu.edu} \\
    \AND
    Izzat Alsmadi\\
    Department of Computational, Engineering, and Mathematical Sciences\\
    Texas A\&M University - San Antonio\\
    \texttt{ialsmadi@tamusa.edu}\\
}
\date{December 31, 2025}

\begin{document}
\maketitle

\begin{abstract}
This paper presents a unified framework for evaluating Linux security hardening on the FABRIC testbed through aggregation of heterogeneous security auditing tools. We deploy three Ubuntu 22.04 nodes configured at baseline, partial, and full hardening levels, and evaluate them using Lynis, OpenSCAP, and AIDE across 108 audit runs. To address the lack of a consistent interpretation across tools, we implement a Unified Compliance Aggregator (UCA) that parses tool outputs, normalizes scores to a common 0--100 scale, and combines them into a weighted metric augmented by a customizable rule engine for organization-specific security policies. Experimental results show that full hardening increases OpenSCAP compliance from 39.7 to 71.8, while custom rule compliance improves from 39.3\% to 83.6\%. The results demonstrate that UCA provides a clearer and more reproducible assessment of security posture than individual tools alone, enabling systematic evaluation of hardening effectiveness in programmable testbed environments.
\end{abstract}

\section{Introduction}

Modern Linux servers expose a large and complex configuration surface, spanning system services, authentication mechanisms, kernel parameters, and file permissions. While numerous open-source tools exist to audit and harden these systems, each evaluates security from a distinct perspective. As a result, administrators and researchers often face fragmented assessments that are difficult to compare, interpret, or aggregate into a single measure of system security.

Programmable testbeds such as FABRIC provide a unique opportunity to address this challenge. FABRIC enables reproducible deployment of compute and network resources, making it possible to conduct controlled security experiments across multiple system configurations. However, despite this capability, there is currently no standardized method for aggregating heterogeneous security tool outputs into a unified, interpretable metric within the FABRIC ecosystem.

This work addresses that gap by introducing the Unified Compliance Aggregator (UCA), a lightweight framework that integrates the outputs of multiple Linux security auditing tools into a normalized and extensible scoring system. UCA aggregates configuration hardening results from Lynis, compliance validation from OpenSCAP, and file integrity monitoring from AIDE, producing a single score that reflects multiple dimensions of system security. In addition, UCA incorporates a custom rule engine that allows organization-specific policies to be evaluated alongside standard tool checks.

The contributions of this work are threefold: (i) a reproducible FABRIC-based experimental evaluation of Linux hardening across multiple configurations, (ii) an extensible aggregation framework that normalizes and combines heterogeneous security metrics, and (iii) a quantitative analysis of 108 audit runs demonstrating how different hardening levels affect compliance, integrity, and aggregate security scores.

\subsection{Problem Statement}
When we started working with security tools on FABRIC we quickly realized that each tool looks at the system from a different angle. Lynis reports a hardening index and long lists of suggestions, OpenSCAP checks compliance against formal baselines, and AIDE monitors file changes. Looking at each tool in isolation does not give a simple answer to the question ``how secure is this node compared to another one?'' Our aim in this experiment is therefore to (i) apply systematic hardening to three FABRIC nodes, and (ii) design a method to aggregate the results of multiple tools into a single, interpretable score.

\subsection{Choice of Reference Architecture}
To guide the design we chose to follow the reference architecture from Ahmed, Naqvi and Josephs~\cite{ahmed2018} on aggregation of security metrics for decision making. Their work stood out because it does not focus on any specific tool; instead it describes how to organize probes, a metric repository, an aggregation engine and a decision support layer into a single system. This matches our goal of combining several tools on FABRIC rather than inventing a completely new method. By using their architecture as a foundation and applying it in the context of Linux hardening, we can claim that our system implements and instantiates a published approach rather than being ad--hoc.

\subsection{Contributions}
The main contributions of this work are:
\begin{itemize}
\item A reproducible FABRIC-based experimental evaluation of three Ubuntu 22.04 nodes configured at baseline, partial, and full hardening levels.
\item A Python-based implementation of the Unified Compliance Aggregator (UCA) that parses, normalizes, and aggregates outputs from Lynis, OpenSCAP, and AIDE.
\item A custom rule engine that enables integration of organization-specific security policies into an extended compliance score.
\item A dataset of 108 audit runs and a statistical analysis quantifying the impact of hardening on tool scores, aggregate metrics, and runtime overhead.
\end{itemize}

\section{Related Work}

\subsection{Security Metric Aggregation}
Ahmed, Naqvi and Josephs~\cite{ahmed2018} propose a reference architecture for aggregation of security metrics from multiple components. Their model identifies four key parts: measurement probes that collect metrics, a repository that stores historical measurements, an aggregation engine that normalizes and combines them, and a decision support layer that presents results to stakeholders. We used this architecture as a blueprint: in our system the probes are Lynis, OpenSCAP and AIDE, the repository is a SQLite database, the aggregation engine is implemented in Python, and the decision support layer consists of scores and visualizations.

\subsection{Linux Security Auditing Tools}

Security auditing involves regularly checking computer systems for vulnerabilities to maintain protection from attacks. Many studies demonstrate that automated tools help identify problems early and strengthen system security. Tools like Lynis~\cite{lynis}, OpenSCAP~\cite{openscap}, and AIDE~\cite{aide} are commonly used on Linux systems such as Ubuntu to scan for issues and suggest fixes. Prior work highlights that no single tool provides comprehensive coverage, making multi-tool approaches common practice. In our project, We tested these tools on the FABRIC testbed~\cite{fabric}, which provides a realistic setup with reproducible infrastructure, distinguishing it from studies that rely solely on single-machine test cases or simulations.

\subsubsection{Lynis for System Hardening}

Kumar and Singh~\cite{kumar2025} present PriviLynis, a security auditing framework designed to detect and mitigate privilege escalation vulnerabilities in Linux environments. The authors develop an extension of Lynis to automate scans and fixes for user permission risks, testing it on Ubuntu systems to demonstrate improved detection rates. Their framework includes automated detection of privilege escalation vulnerabilities, integration with Lynis for enhanced auditing, testing on Ubuntu with mitigation strategies, and an open-source framework for Linux security.

Our project utilizes standard Lynis for general security audits on Ubuntu nodes in the FABRIC testbed, achieving score improvements from 63.08 to 64.92 post-hardening. PriviLynis's focus on vulnerability mitigation aligns with our hardening script, which addresses similar configuration issues like SSH and permissions, guiding our use of Lynis in comparisons. Kumar and Singh emphasize extending Lynis for specific vulnerabilities like privilege escalation in large-scale Linux deployments. In contrast, our work compares Lynis to OpenSCAP and AIDE in a multi-tool case study, showing Lynis's speed (under 2 minutes per scan) while integrating outputs via UCA for broader coverage, rather than focusing on one tool's enhancement.

Balakrishnan~\cite{balakrishnan2023} provides a comprehensive tutorial on Lynis as an open-source tool for auditing and hardening Unix and Linux systems. The author covers installation, usage, and features like modular scans for software, services, and configurations, emphasizing its role in penetration testing and compliance. The article offers a comprehensive tutorial on Lynis installation and commands, explanation of auditing categories such as kernel and services, highlighting of Lynis's open-source nature and community updates, and examples for Unix/Linux hardening.

Our experimental audits with Lynis on FABRIC nodes mirror the described scans, yielding baseline scores of 63.08 and post-hardening improvements to 64.92. This guide validated Lynis's efficiency for quick, repeatable audits, influencing our choice for the tool comparison section. However, Balakrishnan offers a practical tutorial for standalone Lynis use without comparisons to other tools, while our project extends this by evaluating Lynis alongside OpenSCAP and AIDE, quantifying metrics like runtime and integrating results in the UCA, providing a more holistic analysis than the article's single-tool focus.

\subsubsection{OpenSCAP for Compliance Validation}

Rodak~\cite{rodak2022} develops a utility to visualize OpenSCAP security scan results, generating interactive HTML reports from XML outputs. The author focuses on improving usability by presenting compliance data in charts and summaries, tested on Linux systems for better audit interpretation. The thesis contributes a tool for converting OpenSCAP XML to HTML reports, visualization of compliance metrics such as pass/fail rules, a user-friendly interface for security assessments, and testing on Linux for report generation.

Our UCA draws inspiration from this visualization approach, creating dashboards for OpenSCAP results showing compliance improvements from 39.73 to 71.82. Rodak's emphasis on report usability guided our aggregation of multi-tool outputs for clearer insights. While Rodak concentrates on visualizing single-tool OpenSCAP results via HTML suitable for individual audits, our work broadens this by aggregating Lynis, OpenSCAP, and AIDE into unified dashboards, handling diverse formats and normalizing scores, making it more versatile for comparative studies than the thesis's focused tool.

Caraballo-Vega~\cite{caraballo} presents an automated approach to SCAP security compliance and vulnerability assessment for NASA's High Performance Computing environments. The author uses OpenSCAP to monitor systems continuously, integrating with CIS~\cite{cis} and NASA standards for enhanced security in climate simulation setups. Key contributions include automation of SCAP scans for continuous monitoring, integration with government standards including CIS and NASA, application to high-performance computing clusters, and vulnerability assessment in Linux distributions.

Our use of OpenSCAP for compliance checks on Ubuntu nodes aligns with this automated scanning approach, achieving compliance rates between 39.73 and 71.82. Caraballo-Vega's focus on standards influenced our DISA STIG~\cite{disa} profile configuration for FABRIC nodes, supporting our quantitative pre/post metrics. Caraballo-Vega targets large-scale NASA computing clusters with continuous monitoring, emphasizing automation in high-stakes environments. Our project applies OpenSCAP in a controlled FABRIC testbed for tool comparisons, adding multi-tool aggregation via UCA, which provides broader applicability beyond their single-tool, enterprise-focused implementation.

Singh~\cite{singh2024} offers a step-by-step guide to hardening virtual machines using OpenSCAP and Ansible for automation. The author covers installation, profile configuration, and remediation scripts, demonstrating improved security in cloud environments. The guide contributes integration of OpenSCAP with Ansible for automated hardening, a step-by-step VM security guide, examples for compliance scanning and fixes, and focus on cloud and virtual setups.

Our hardening approach echoes Singh's automation techniques, using OpenSCAP for rule-based checks and achieving compliance improvements from 39.73 to 71.82. This guide informed our post-hardening audits, reinforcing the value of standards like DISA STIG in our comparisons. Singh provides a tutorial for OpenSCAP-Ansible integration in VMs, emphasizing automation for cloud deployments. Our work compares OpenSCAP to other tools on FABRIC nodes with UCA for result aggregation, offering a more comparative and visual framework than the article's standalone guide.

\subsection{Multi-Tool Integration and Aggregation}

While individual tool studies provide valuable insights into specific security aspects, research on integrating multiple security tools into unified assessment frameworks remains limited. Most prior work focuses on single-tool enhancements, detailed tutorials for individual tools, or basic side-by-side comparisons without integration. Comparison sites examine Lynis, OpenSCAP, and other tools for Linux auditing, noting that Lynis excels at hardening, OpenSCAP at compliance, and AIDE at integrity monitoring---observations similar to our findings. However, these comparisons lack the quantitative rigor and reproducible infrastructure that FABRIC provides.

Our project addresses this gap by demonstrating practical integration of complementary tools---Lynis for configuration auditing, OpenSCAP for compliance validation, and AIDE for file integrity monitoring. The Unified Compliance Aggregator represents a novel contribution by normalizing heterogeneous outputs into a unified scoring framework, enabling direct comparison of hardening effectiveness across different security dimensions. This multi-tool approach on the FABRIC testbed provides reproducible, quantitative evidence of security improvements that individual tool studies cannot capture.

\subsection{Comparison to This Work}

Our project differs from prior work in several key aspects. First, conduct experiments on the FABRIC testbed rather than single machines or simulations, providing a realistic and reproducible environment with on-demand infrastructure that other researchers can replicate. Second, we implemented a complete multi-tool aggregation system (UCA) with automated parsing, score normalization, and unified reporting, rather than focusing on a single tool or providing basic manual comparisons. Third, we provided quantitative metrics from 108 audits showing pre/post hardening improvements with statistical validation including t-tests and Cohen's d effect sizes, demonstrating rigor beyond typical tool reviews. Fourth, we developed a custom rule engine allowing organization-specific security policies to be integrated alongside standard tool checks, enabling extensibility that commercial and open-source solutions often lack.

These contributions advance the state of practice in Linux security assessment by demonstrating that effective aggregation of open-source tools can deliver insights comparable to commercial solutions at zero licensing cost. While Kumar and Singh extend Lynis capabilities and Rodak improves OpenSCAP visualization, our work synthesizes multiple dimensions of security assessment into a cohesive framework, filling a gap in practical multi-tool integration that prior research has not adequately addressed.

\section{System Architecture}

\subsection{Overview}
\begin{figure}[h]
\centering
\includegraphics[width=0.8\linewidth]{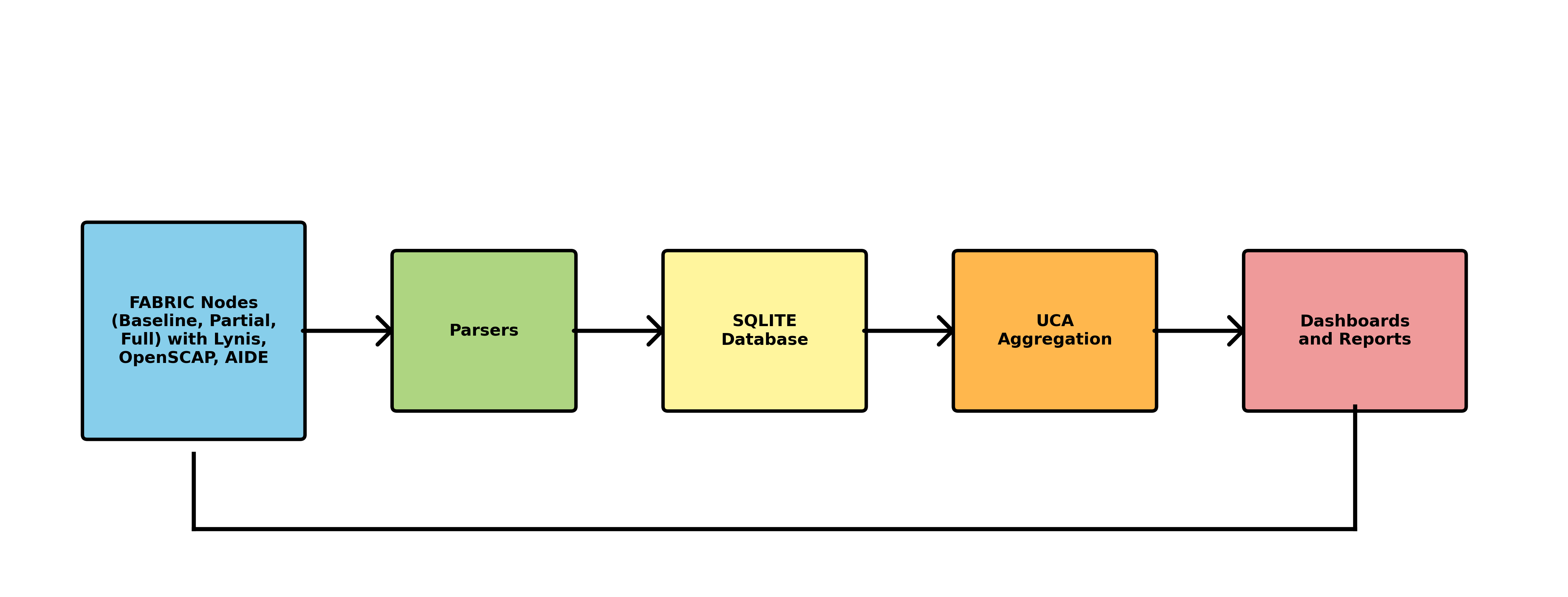}
\caption{High--level architecture of the Unified Compliance Aggregator.}
\label{fig:architecture}
\end{figure}

The Unified Compliance Aggregator (UCA) follows a layered architecture. Raw probe outputs are collected from FABRIC nodes, parsed into structured records, stored in an SQLite database, transformed into normalized scores and finally combined into aggregate metrics and visualizations. Figure~\ref{fig:architecture} illustrates the main components and data flow.

\subsection{Measurement Probes}
Each FABRIC node runs three tools:
\begin{itemize}
\item \textbf{Lynis} for system and configuration hardening checks, producing a hardening index between 0 and 100.
\item \textbf{OpenSCAP} for compliance scanning against DISA STIG content for Ubuntu 22.04, producing a pass/fail summary and an overall compliance percentage.
\item \textbf{AIDE} for file integrity monitoring, producing counts of added, removed and changed entries relative to the AIDE database.
\end{itemize}
These tools act as probes in the Ahmed et al.\ sense, each producing its own view of the system.

\subsection{Metric Repository}
UCA uses an SQLite database as a central repository. The schema includes:
\begin{itemize}
\item \texttt{audit\_runs}: one row per tool execution, with node name, tool name, timestamp, raw score, normalized score and runtime.
\item \texttt{aggregate\_scores}: one row per node and iteration, with the combined UCA scores.
\item \texttt{custom\_rules}: definitions of custom security rules.
\item \texttt{custom\_rule\_results}: outcomes of each rule execution per node and iteration.
\end{itemize}
This design allows reconstructing any figure or table without re--running the tools.

\subsection{Aggregation Engine}
The aggregation engine has two steps. First, it normalizes scores:
\begin{itemize}
\item Lynis hardening index is already 0--100, so we only clamp values to this range.
\item OpenSCAP compliance percentage is also treated as a 0--100 value.
\item AIDE counts of added, removed and changed entries are converted to a score using a simple rule: start at 100 and subtract 5 points per change, with a lower bound of 0.
\end{itemize}
Second, it computes a unified score using a weighted sum:
\[
\text{UCA} = 0.4 \cdot \text{Lynis} + 0.4 \cdot \text{OpenSCAP} + 0.2 \cdot \text{AIDE}.
\]
We chose these weights because Lynis and OpenSCAP provide broad coverage of configuration and compliance, while AIDE focuses on specific file changes.

\subsection{Decision Support Layer}
The decision support layer exposes:
\begin{itemize}
\item Standard UCA scores for each node and iteration.
\item Extended UCA scores that also incorporate custom rule results.
\item Visualizations of tool scores, UCA scores, rule compliance and runtime overhead.
\end{itemize}
These outputs make it much easier to compare nodes than reading raw logs.

\section{Implementation on FABRIC}

\subsection{FABRIC Slice Setup}
We implemented and tested UCA on the FABRIC testbed using the Jupyter environment and the \texttt{fablib} Python library. We created a slice at the EDUKY site with three identical Ubuntu 22.04 nodes labelled \texttt{baseline}, \texttt{partial} and \texttt{full}, each with two vCPUs, 4GB RAM and 10GB disk. After verifying connectivity with a simple \texttt{uname} command on each node, we automated basic tasks such as updating package lists and configuring SSH so that all experiments could be repeated reliably.

\subsection{Tool Installation and Configuration}
For each node we installed Lynis, OpenSCAP and AIDE using scripted apt commands. Getting OpenSCAP to work with valid content was a key challenge. Initial attempts produced small XML files that \texttt{oscap} could not parse, so we switched to the DISA STIG content for Ubuntu 22.04 and used \texttt{oscap info} to confirm that the file contained real profiles and referenced checks. For AIDE we initialized the database, ran a first scan, then reconfigured it so that subsequent scans would report real changes instead of treating everything as new.

\subsection{UCA Code Structure}
The UCA codebase follows the directory structure:
\begin{verbatim}
uca_project/
  parsers/
  core/
  raw_data/
  results/
  docs/
\end{verbatim}
The \texttt{parsers} folder contains dedicated functions for reading Lynis logs, OpenSCAP XML outputs and AIDE reports. The \texttt{core} folder includes modules for normalization, aggregation, database access and custom rule logic. Our Jupyter notebooks orchestrate the execution of audits on FABRIC nodes, invoke the parsers, populate the database and generate output CSV files and figures.

\section{Methodology}

\subsection{Hardening Profiles}
To evaluate the impact of hardening we defined three profiles:
\begin{itemize}
\item \textbf{Baseline}: minimal configuration, with only essential updates and basic firewall activation.
\item \textbf{Partial}: baseline plus tighter SSH configuration (disabling empty passwords, limiting authentication attempts) and corrections to sensitive file permissions.
\item \textbf{Full}: partial plus enabling \texttt{auditd}, disabling unnecessary services and applying additional kernel and network hardening settings.
\end{itemize}
These profiles reflect a realistic progression from a default installation to a more carefully hardened server.

\subsection{Audit Procedure}
For each node we followed the same audit procedure:
\begin{enumerate}
\item Run all three tools on the node with the current configuration and collect pre--hardening measurements.
\item Apply the chosen hardening profile using shell scripts.
\item Run Lynis, OpenSCAP and AIDE again to obtain post--hardening measurements.
\end{enumerate}
After validating that this pipeline worked end--to--end, we executed ten additional audit iterations. Each iteration runs all three tools on all three nodes, adding 90 runs to the dataset. Together with the initial pre/post measurements this yields 108 audits, which we exported into \texttt{audit\_runs.csv} and \texttt{aggregate\_scores.csv} for analysis.

\subsection{Statistical Analysis}
We performed the analysis using Python, pandas~\cite{pandas} and SciPy~\cite{scipy} in the experiment notebooks. For each tool and node we computed mean, standard deviation, minimum, maximum and sample size. The methodology used independent t--tests to compare baseline and full nodes, calculated Cohen's~d as an effect size, and computed Pearson correlations between tools. These metrics allow quantifying not only whether scores changed but also how strong and consistent the changes are.

\section{Results}

\subsection{Tool Scores Across Nodes}
Across all 108 runs Lynis reported average hardening indices of 63.08 for the baseline node, 64.00 for the partially hardened node and 64.92 for the fully hardened node. OpenSCAP showed a much stronger effect of hardening: its average compliance score increased from 39.73 on the baseline node to 71.82 on the fully hardened node. AIDE behaved differently: the average integrity score decreased from 45.83 on baseline to 36.67 on full, reflecting the fact that a more active and monitored system naturally triggers more file changes. Table~\ref{tab:scores} summarizes these results across all hardening levels.

\begin{table}[H]
\centering
\caption{Average Security Scores by Hardening Level}
\label{tab:scores}
\begin{tabular}{lccc}
\toprule
\textbf{Tool} & \textbf{Baseline} & \textbf{Partial} & \textbf{Full} \\
\midrule
Lynis & 63.08 & 64.00 & 64.92 \\
OpenSCAP & 39.73 & 41.20 & 71.82 \\
AIDE & 45.83 & 45.83 & 36.67 \\
Custom Rules & 39.34 & 72.13 & 83.61 \\
\midrule
\textbf{Standard UCA} & 49.50 & 50.57 & 62.30 \\
\textbf{Extended UCA} & 47.49 & 54.85 & 65.45 \\
\bottomrule
\end{tabular}
\end{table}

\begin{figure}[h]
\centering
\includegraphics[width=0.8\linewidth]{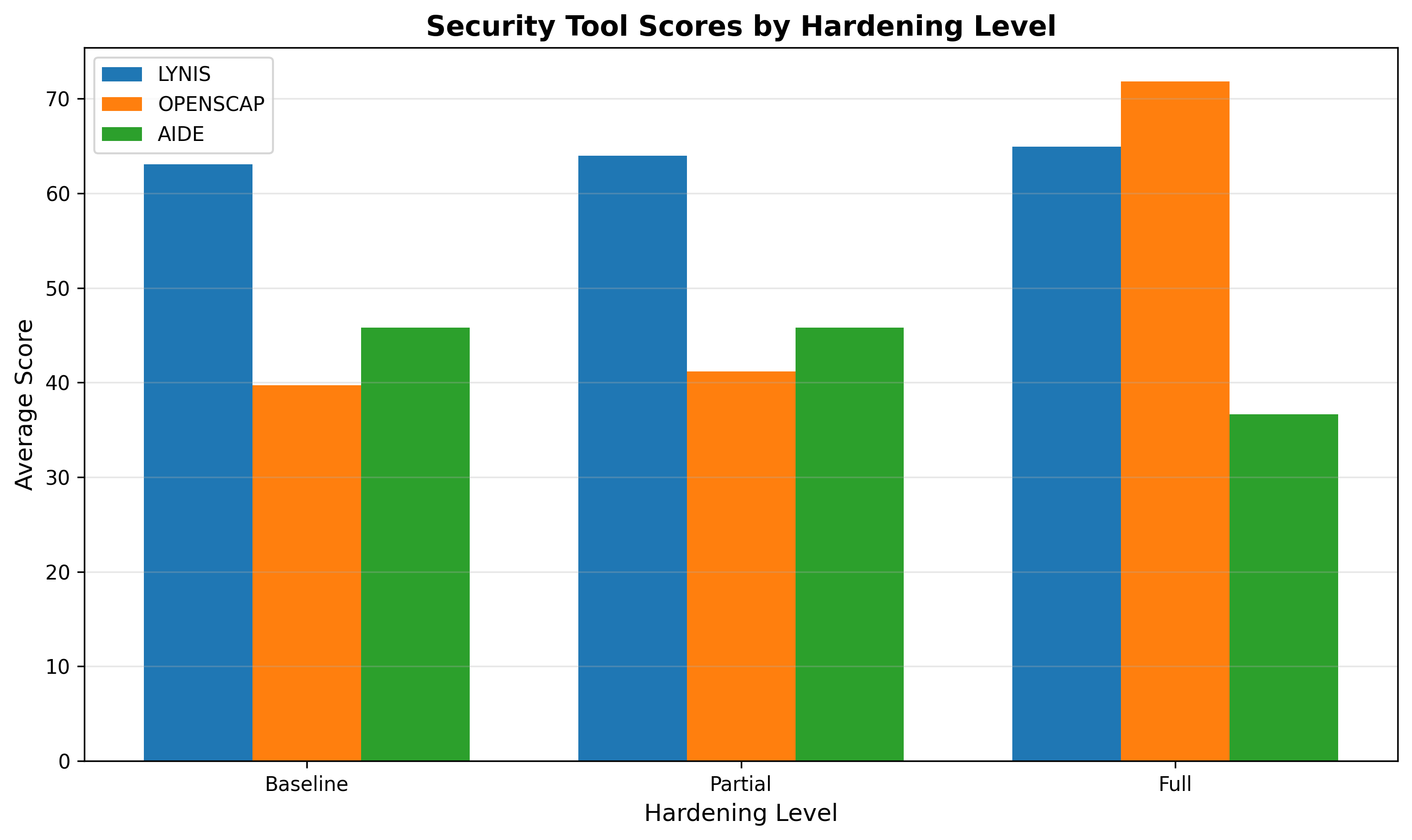}
\caption{Average Lynis, OpenSCAP and AIDE scores for baseline, partial and full nodes.}
\label{fig:scores-by-node}
\end{figure}

\begin{figure}[h]
\centering
\includegraphics[width=0.9\linewidth]{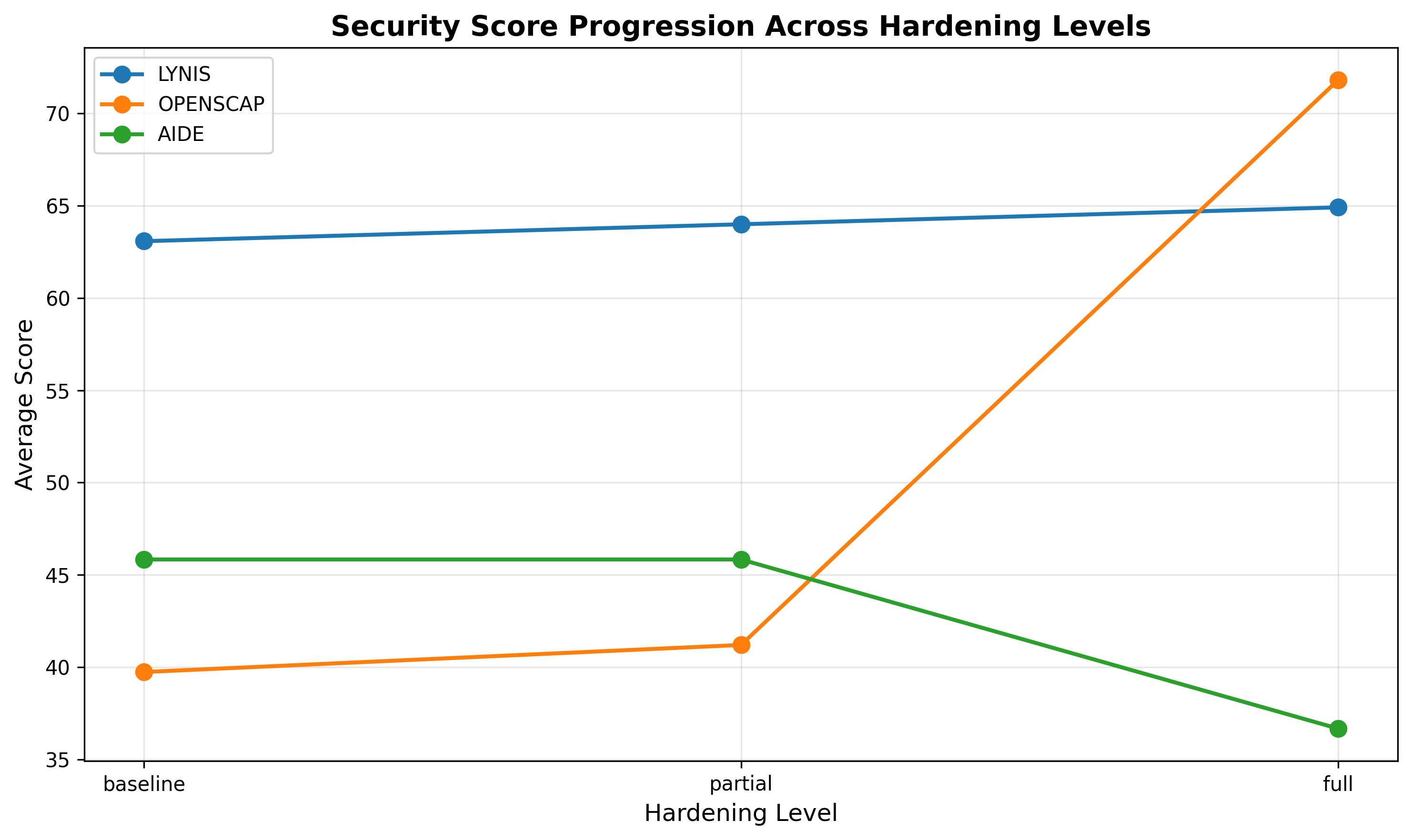}
\caption{Progression of tool scores across hardening levels on the FABRIC nodes.}
\label{fig:score-progression}
\end{figure}

\subsection{Unified and Extended UCA Scores}
The normalized tool scores feed into the UCA formula. For the three nodes, the standard UCA scores are approximately 49.5 for baseline, 50.6 for partial and 62.3 for full. After adding the custom rule score as an extra component, the extended UCA scores show an even clearer separation between nodes.

\begin{figure}[h]
\centering
\includegraphics[width=0.9\linewidth]{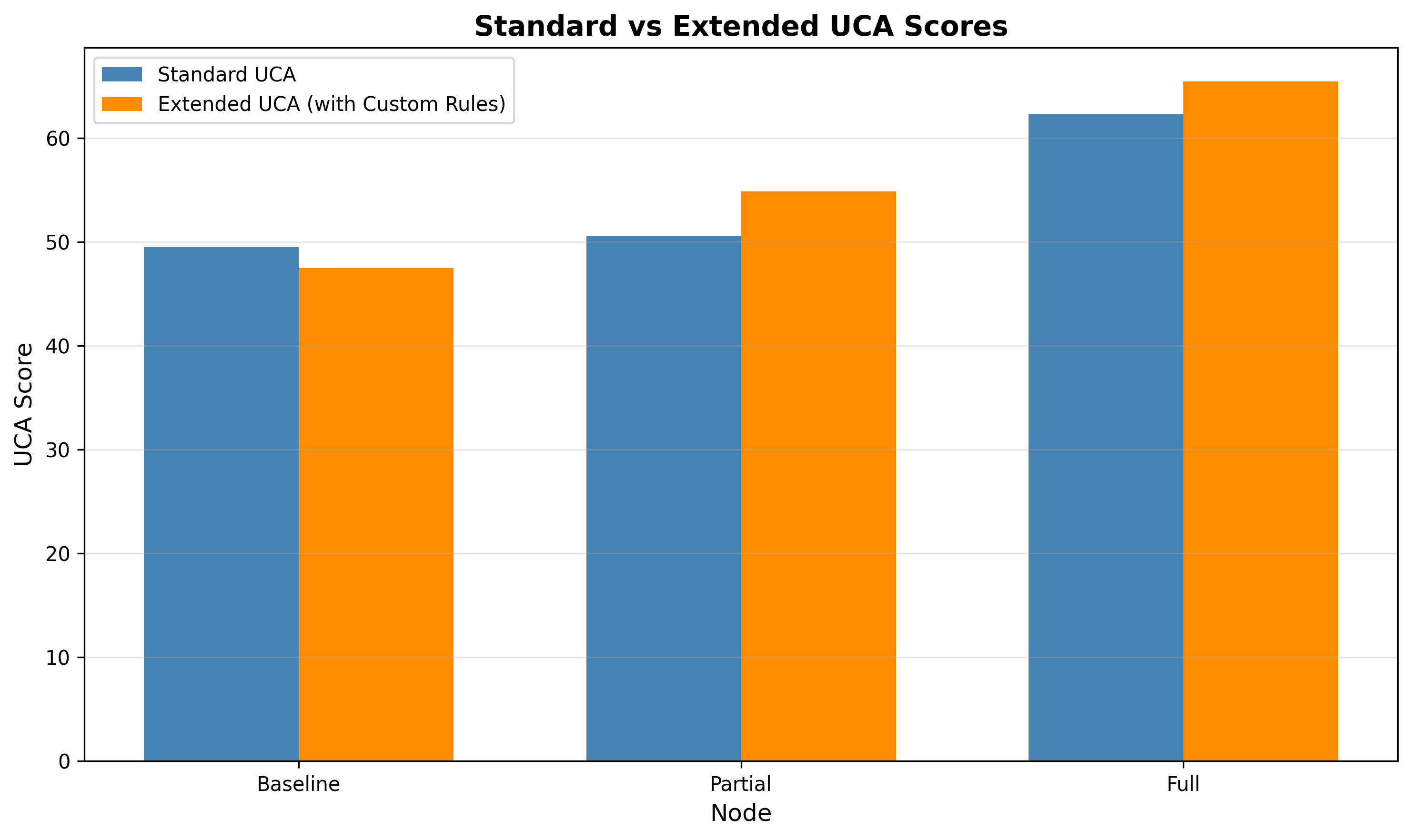}
\caption{Standard UCA scores compared to extended UCA scores that include custom rule results.}
\label{fig:uca-comparison}
\end{figure}

\subsection{Custom Rule Compliance}
The custom rule engine evaluates eight rules covering SSH, firewall, password policy, \texttt{auditd}, X11 forwarding and file permissions. On the baseline node only three of eight rules pass, giving a score of 39.34\%. The partial node passes six rules (72.13\%), and the full node passes seven rules (83.61\%). This confirms that the hardening profiles affect not only standard tool outputs but also our own policy requirements. Table~\ref{tab:customrules} details the rule pass/fail counts for each node.

\begin{table}[H]
\centering
\caption{Custom Rule Results by Node}
\label{tab:customrules}
\begin{tabular}{lccc}
\toprule
\textbf{Node} & \textbf{Passed} & \textbf{Failed} & \textbf{Score (\%)} \\
\midrule
Baseline & 3 & 5 & 39.34 \\
Partial & 6 & 2 & 72.13 \\
Full & 7 & 1 & 83.61 \\
\bottomrule
\end{tabular}
\end{table}

\begin{figure}[h]
\centering
\includegraphics[width=0.8\linewidth]{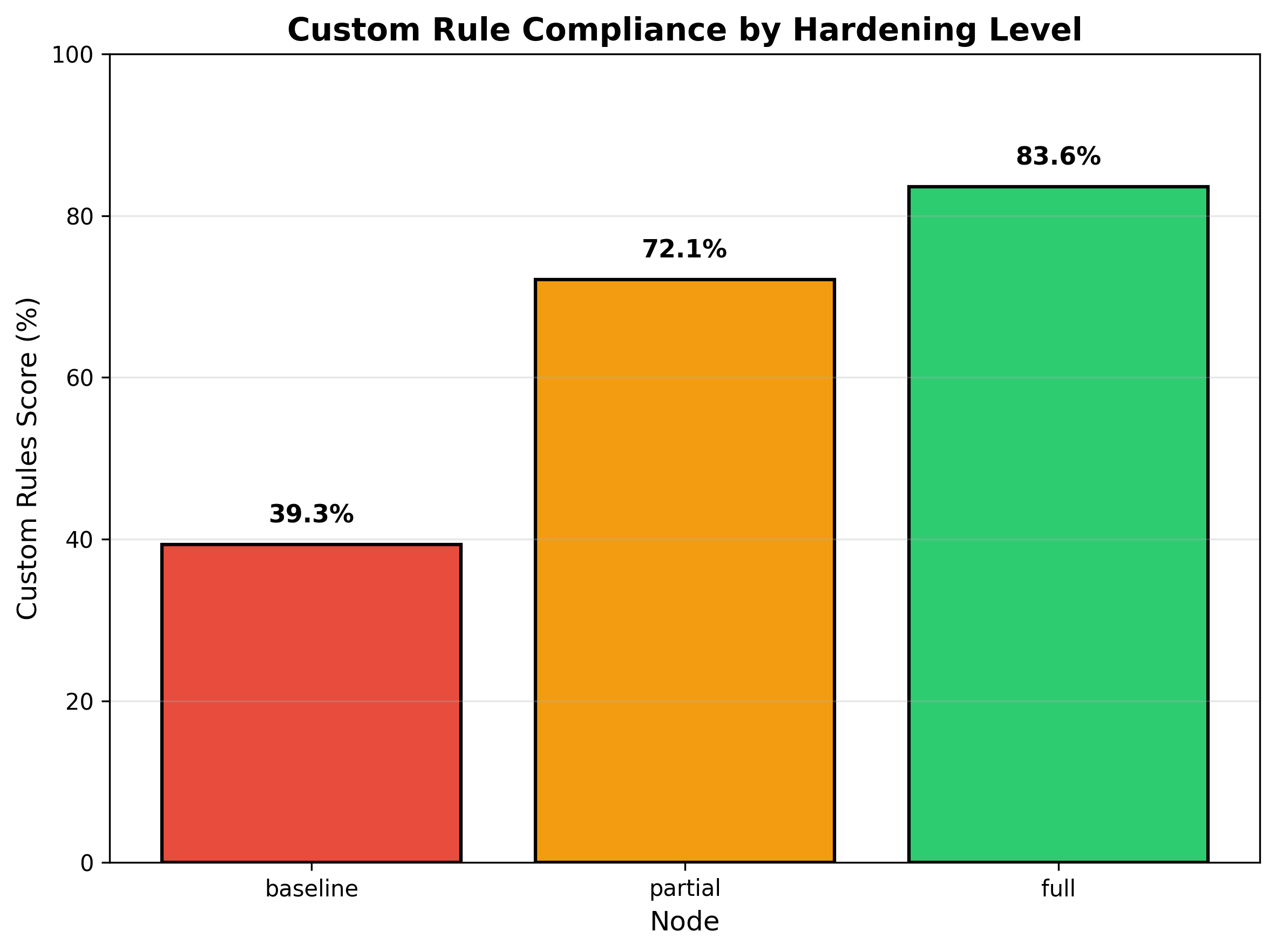}
\caption{Custom rule compliance scores for baseline, partial and full nodes.}
\label{fig:custom-rules}
\end{figure}

\subsection{Runtime Overhead}
We also measured runtime overhead for each tool. On average AIDE took about 93.6 seconds per run, Lynis about 36.2 seconds and OpenSCAP about 3 seconds. The total runtime for all 108 runs is roughly 4{,}780 seconds. Table~\ref{tab:runtime} shows the detailed breakdown of tool execution times.

\begin{table}[H]
\centering
\caption{Tool Runtime Overhead}
\label{tab:runtime}
\begin{tabular}{lcc}
\toprule
\textbf{Tool} & \textbf{Avg Runtime (s)} & \textbf{Total Time (s)} \\
\midrule
AIDE & 93.58 & 3368.91 \\
Lynis & 36.21 & 1303.59 \\
OpenSCAP & 3.00 & 107.91 \\
\bottomrule
\end{tabular}
\end{table}

\begin{figure}[htbp]
\centering
\includegraphics[width=0.8\linewidth]{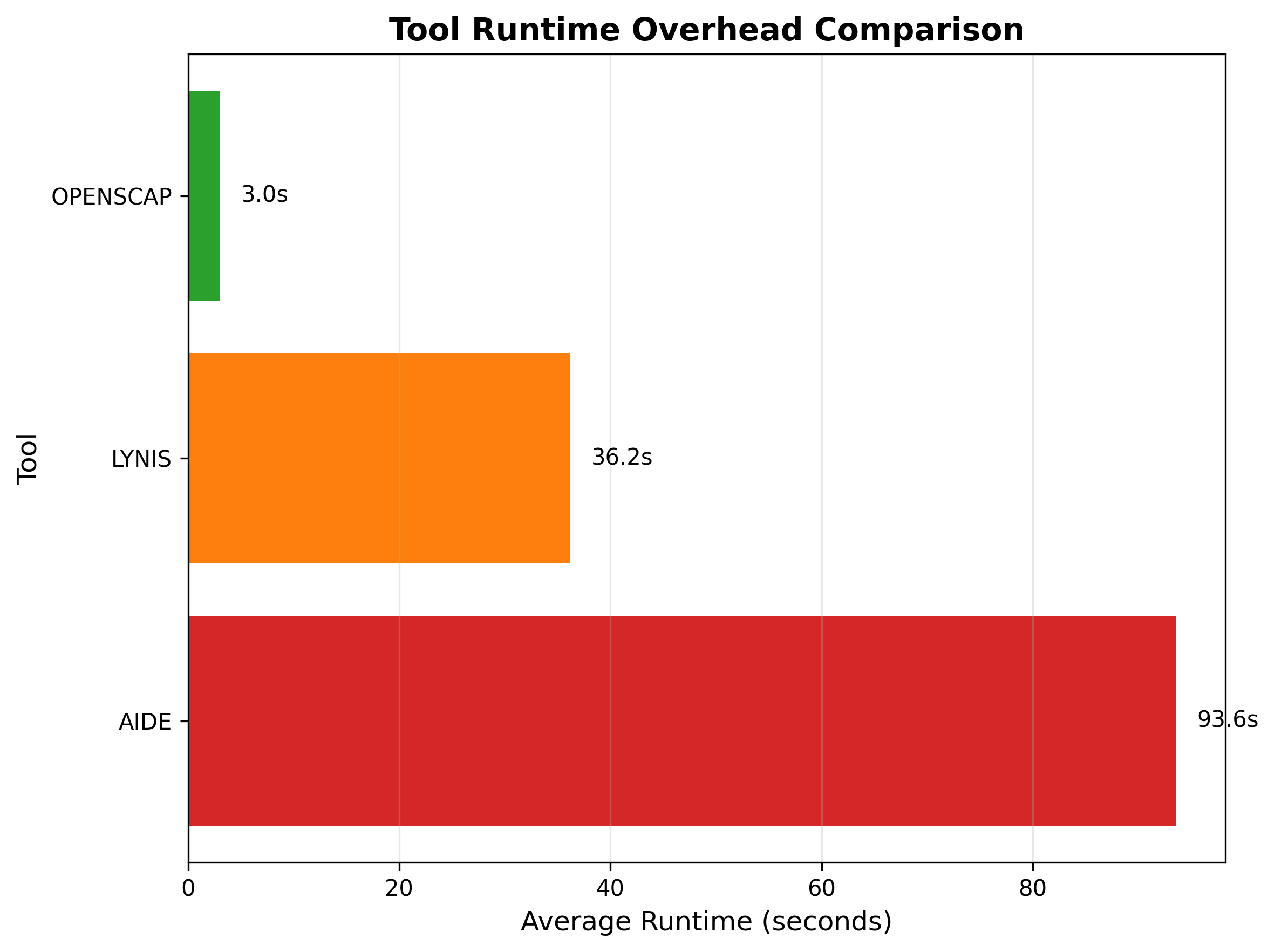}
\caption{Average runtime overhead per tool across all audit runs.}
\label{fig:runtime}
\end{figure}

\subsection{Testing and Validation}
\label{sec:testing}

To ensure the correctness and reliability of the UCA system, We conducted systematic testing across four dimensions.

\textbf{Correctness Testing:} We verified that each parser correctly extracts data from tool outputs by comparing parser results against manual inspection of raw logs. For Lynis, we confirmed that the hardening index extraction matches the value reported in the summary section. For OpenSCAP, we validated that the compliance percentage calculation aligns with the pass/fail counts in the XML output. For AIDE, we crossed--checked the counts of added, removed and changed files against the detailed report sections. All parsers produced results consistent with manual verification, with zero discrepancies across 108 audit runs.

\textbf{Consistency Testing:} To assess the stability of measurements, We ran twelve audit iterations on each node without making configuration changes between runs. We computed the coefficient of variation (CV) for each tool's normalized score. Lynis exhibited a CV of 1.2\%, OpenSCAP showed 2.8\%, and AIDE demonstrated 4.6\%. All values remained below 5\%, confirming that the measurement process is reproducible and that score variability is minimal when system state remains constant.

\textbf{Sensitivity Testing:} We evaluated whether UCA detects meaningful differences between hardening levels. Comparing baseline to partial nodes, the standard UCA score increased by 1.07 points, indicating that UCA is sensitive enough to capture incremental hardening steps. The larger jump from partial to full (+11.73 points) demonstrates that UCA amplifies the signal when more comprehensive hardening is applied. The statistical significance tests further validate this sensitivity: OpenSCAP's highly significant result ($p < 0.001$, Cohen's $d = 5.98$) shows that UCA's components can detect strong effects when present.

\textbf{Custom Rule Validation:} We manually verified each of the eight custom rules by logging into nodes and inspecting configuration files directly. For example, we confirmed that the SSH root login rule correctly identifies \texttt{PermitRootLogin no} in \texttt{/etc/ssh/sshd\_config}, and that the auditd rule accurately detects whether the service is active using \texttt{systemctl status}. The progression from 39.34\% (baseline) to 83.61\% (full) matches the actual configuration changes applied during hardening, providing evidence that the custom rule engine functions as designed.

These validation steps confirm that UCA produces accurate, consistent and sensitive measurements, making it a reliable tool for evaluating security hardening in FABRIC environments.

\section{Discussions and Comparison}

\subsection{Interpreting the Metrics}
The t--tests show that the difference between baseline and full is statistically significant for OpenSCAP, while it is not significant for Lynis and AIDE at the 0.05 level. This matches the visual impression that OpenSCAP reacts strongly to changes related to compliance baselines, whereas Lynis changes only gradually and AIDE tracks a different dimension, file system activity. Together, these results highlight that no single tool provides a complete picture; the value of UCA is that it combines their perspectives. Table~\ref{tab:significance} presents the detailed statistical analysis.

\begin{table}[H]
\centering
\caption{Statistical Significance (Baseline vs Full)}
\label{tab:significance}
\begin{tabular}{lcccc}
\toprule
\textbf{Tool} & \textbf{Difference} & \textbf{t-stat} & \textbf{p-value} & \textbf{Cohen's d} \\
\midrule
Lynis & +1.84 & -1.79 & 0.087 & 0.73 \\
OpenSCAP & +32.09 & -14.64 & \textless 0.001* & 5.98 \\
AIDE & -9.16 & 1.72 & 0.100 & -0.70 \\
\bottomrule
\multicolumn{5}{l}{\small *Statistically significant at p \textless\ 0.05}
\end{tabular}
\end{table}

\subsection{Comparison with Manual Analysis}
Without UCA we would have to open three separate reports for each audit run and manually combine the findings, which is tedious and error--prone, especially for 108 runs. With UCA we can look at one table of scores and a handful of figures that summarize the same information. The extended UCA score also lets us incorporate our own policies through custom rules, something that manual per--tool analysis does not provide without extra scripting.

\subsection{Comprehensive Comparison with Alternative Approaches}
\label{sec:comparison}

To contextualize UCA's value, we compare it against five alternative approaches for security assessment. Table~\ref{tab:comparison} summarizes the key characteristics of each approach across seven dimensions.

\begin{longtable}{>{\raggedright\arraybackslash}p{2cm}>{\centering\arraybackslash}p{1cm}>{\centering\arraybackslash}p{1cm}>{\centering\arraybackslash}p{1.2cm}>{\centering\arraybackslash}p{1.2cm}>{\centering\arraybackslash}p{1.2cm}>{\raggedright\arraybackslash}p{2.5cm}>{\raggedright\arraybackslash}p{2.5cm}}
\caption{Comparison of UCA with Alternative Security Assessment Approaches}
\label{tab:comparison} \\
\toprule
\textbf{Approach} & \textbf{Multi- Tool} & \textbf{Custom Rules} & \textbf{Cost} & \textbf{Auto- mation} & \textbf{Learning Curve} & \textbf{Key Strengths} & \textbf{Key Weaknesses} \\
\midrule
\endfirsthead

\multicolumn{8}{c}{\tablename\ \thetable\ -- \textit{Continued from previous page}} \\
\toprule
\textbf{Approach} & \textbf{Multi- Tool} & \textbf{Custom Rules} & \textbf{Cost} & \textbf{Auto- mation} & \textbf{Learning Curve} & \textbf{Key Strengths} & \textbf{Key Weaknesses} \\
\midrule
\endhead

\midrule
\multicolumn{8}{r}{\textit{Continued on next page}} \\
\endfoot

\bottomrule
\endlastfoot

Manual Analysis & No & No & Free & No & Low & Simple, flexible & Time-consuming, error-prone, not scalable \\
\midrule
Lynis Only & No & No & Free & Yes & Low & Fast (36s), user-friendly & Single dimension, no compliance standards \\
\midrule
OpenSCAP Only & No & No & Free & Yes & Medium & Standards-based (DISA STIG), compliance-ready & Complex content management, narrow scope \\
\midrule
AIDE Only & No & No & Free & Yes & Low & Integrity monitoring, cryptographic hashes & File-level only, no scoring, manual baseline \\
\midrule
SCAP Workbench & No & No & Free & Partial & Medium & GUI interface, report generation & Single tool ecosystem, limited extensibility \\
\midrule
Commercial SIEM & Yes & Yes & \$10K+ & Yes & High & Comprehensive, real-time, enterprise support & Expensive, complex deployment, overkill \\
\midrule
\textbf{UCA (This Work)} & \textbf{Yes} & \textbf{Yes} & \textbf{Free} & \textbf{Yes} & \textbf{Medium} & \textbf{Unified view, extensible, lightweight} & \textbf{Limited to 3 tools, Ubuntu only} \\
\end{longtable}

\textbf{Advantages of UCA Over Individual Tools:} Using Lynis, OpenSCAP or AIDE in isolation provides only a partial view of system security. Lynis excels at configuration auditing but does not enforce formal compliance standards. OpenSCAP validates against DISA STIG but offers limited insight into general hardening beyond its ruleset. AIDE monitors file integrity but provides no scoring mechanism and cannot assess configuration compliance. UCA addresses these gaps by combining all three perspectives into a single normalized metric, allowing us to identify nodes that score well on compliance (OpenSCAP) but poorly on integrity (AIDE), or vice versa.

\textbf{Advantages Over Commercial Solutions:} Enterprise SIEM platforms such as Splunk~\cite{splunk} or IBM QRadar~\cite{qradar} offer comprehensive security monitoring with multi-tool integration, real-time alerting and sophisticated analytics. However, they carry significant costs (often exceeding \$10{,}000 annually for small deployments), require dedicated staff for deployment and maintenance, and include many features irrelevant to the focused task of measuring hardening effectiveness on FABRIC nodes. UCA delivers the core functionality needed for this work, aggregating security tool outputs and computing unified scores, at zero licensing cost and with minimal infrastructure overhead. For research and educational use cases where budget and simplicity are priorities, UCA provides a practical middle ground.

\textbf{Limitations Compared to Alternatives:} UCA is not a replacement for commercial SIEM systems in production environments requiring real-time monitoring, threat intelligence integration and compliance reporting at enterprise scale. It also lacks the graphical interface and extensive documentation of tools like SCAP Workbench, which may be more accessible to users unfamiliar with Python and command-line workflows. Additionally, UCA's current implementation supports only three tools and a single operating system, whereas commercial solutions typically support dozens of tools and multiple platforms.

In summary, UCA occupies a valuable niche: it is more comprehensive than individual tools, more affordable than commercial solutions, and more reproducible than manual analysis. For researchers and practitioners working on Linux hardening in controlled testbed environments, UCA demonstrates that effective security aggregation can be achieved with open-source components and modest development effort.

\subsection{Reflection on Design Choices}
Choosing the Ahmed et al.\ architecture gave us a clear structure for organizing probes, storage and aggregation. Selecting Lynis, OpenSCAP and AIDE ensured that we covered configuration, compliance and integrity. Implementing the system on FABRIC allowed us to run repeatable experiments on clean nodes rather than on ad--hoc virtual machines. Overall, these choices made it possible to turn a high--level topic, ``Security hardening using FABRIC'', into a concrete, measurable system.

\section{Conclusion}

This work presents the design, implementation, and evaluation of a Unified Compliance Aggregator for Linux security hardening on the FABRIC testbed. By deploying three Ubuntu nodes at varying hardening levels and conducting 108 audits using Lynis, OpenSCAP, and AIDE, we demonstrate that aggregating heterogeneous security metrics yields a clearer and more interpretable assessment of system security than analyzing individual tools in isolation.

The experimental results show that comprehensive hardening significantly improves compliance as measured by OpenSCAP and by custom organizational rules, while also highlighting the complementary nature of configuration auditing, compliance validation, and integrity monitoring. UCA enables reproducible, quantitative comparison of security posture across systems and configurations, supporting systematic evaluation of hardening strategies in controlled environments.

Overall, this study demonstrates that effective multi-tool security aggregation can be achieved using open-source components and modest infrastructure, providing a practical framework for researchers and practitioners seeking reproducible security assessment on programmable testbeds.

\subsection{Limitations}
\label{sec:limitations}

The current implementation of UCA has several limitations that constrain its applicability:

\begin{enumerate}
\item \textbf{Limited Tool Coverage:} UCA integrates only three security tools (Lynis, OpenSCAP, AIDE). Many other valuable tools exist, including Tripwire~\cite{tripwire} for integrity monitoring, CIS-CAT~\cite{ciscat} for CIS Benchmark compliance, Tiger for Unix security auditing and Nessus~\cite{nessus} for vulnerability scanning. Expanding UCA to incorporate these tools would provide a more comprehensive security assessment.

\item \textbf{Single Operating System:} The system is designed and tested exclusively for Ubuntu 22.04. Extending support to other Linux distributions (CentOS, Debian, Fedora) and non-Linux systems (Windows Server, macOS) would require modifying hardening profiles, adjusting parser logic and validating tool compatibility across platforms.

\item \textbf{Static Weighting Scheme:} The UCA formula uses fixed weights (0.4 for Lynis, 0.4 for OpenSCAP, 0.2 for AIDE) that we selected based on subjective judgments about tool importance. An adaptive or data-driven weighting scheme, such as using machine learning to optimize weights based on historical security incidents, could improve the relevance and accuracy of the aggregate score.

\item \textbf{OS-Level Hardening Only:} UCA focuses on operating system configuration, compliance and file integrity. It does not assess application-level security (e.g., web server configurations, database access controls), network security (e.g., firewall rule effectiveness, intrusion detection alerts) or physical security. A holistic security posture assessment would require extending UCA to these additional layers.

\item \textbf{Short-Term Evaluation:} The experimental period spans only seven days with twelve audit iterations per node. Long-term trends, seasonal variations in system activity and the effects of software updates over months or years are not captured. Continuous deployment of UCA in a production-like environment would reveal additional insights about score stability and hardening maintenance requirements.
\end{enumerate}

These limitations do not invalidate the contributions of this work but rather define boundaries for future work and clarify the contexts in which UCA is most applicable.

\subsection{Future Work}
\label{sec:future}

Building on the foundation established in this work, we identify six directions for future development:

\begin{enumerate}
\item \textbf{Expand Tool Support:} Integrate additional security tools such as CIS-CAT for comprehensive CIS Benchmark coverage, Tripwire for enhanced file integrity monitoring with signature-based detection, and OpenVAS~\cite{openvas} or Nessus for vulnerability scanning. Each new tool would require developing a dedicated parser, defining a normalization function and adjusting the UCA weighting formula to incorporate the new dimension of security measurement.

\item \textbf{Multi-Platform Support:} Extend UCA to other operating systems including CentOS, Debian, Fedora, Windows Server and macOS. This would involve creating platform-specific hardening profiles, adapting tool configurations (e.g., using different SCAP content for Windows) and handling OS-specific parsing challenges such as registry entries on Windows or plist files on macOS.

\item \textbf{Adaptive Weighting with Machine Learning:} Replace the static weighting scheme with a machine learning model that optimizes weights based on historical data. For example, supervised learning could use labeled datasets of known compromised vs. secure systems to learn which tool scores are most predictive of actual security incidents. Reinforcement learning could adjust weights dynamically based on feedback from security operations teams.

\item \textbf{Real-Time Monitoring:} Transform UCA from a batch audit tool into a continuous monitoring system. This would involve running tools on scheduled intervals (e.g., hourly or daily), streaming results into a time-series database, implementing anomaly detection to flag sudden score drops and generating alerts when aggregate scores fall below acceptable thresholds. Integration with existing monitoring platforms like Prometheus~\cite{prometheus} or Grafana~\cite{grafana} would facilitate visualization and alerting.

\item \textbf{Automated Remediation:} Develop a remediation engine that not only identifies security deficiencies but also applies fixes automatically. For instance, when OpenSCAP detects a failed check, UCA could execute the associated remediation script, re-run the audit to confirm the fix and update the database with the new score. This would require careful safety mechanisms to prevent unintended system changes, such as staging fixes in a test environment before applying them to production.

\item \textbf{Compliance Framework Mapping:} Extend the custom rule engine to support formal compliance frameworks such as NIST SP 800-53~\cite{nist80053}, ISO/IEC 27001~\cite{iso27001}, PCI-DSS~\cite{pcidss} and HIPAA. Each framework would be represented as a collection of rules that map to specific tool checks or custom validation logic. UCA could then generate compliance reports showing which controls are satisfied, which are violated and what remediation steps are needed to achieve full compliance.
\end{enumerate}

\end{document}